\documentclass{article}

\usepackage[inference]{semantic}
\usepackage{amsmath}\allowdisplaybreaks
\usepackage[ntheorem]{mathtools}
\usepackage{hyperref}
\usepackage{comment}
\usepackage[amsmath,hyperref,amsthm]{ntheorem}
\usepackage{thmtools}
\usepackage{nameref}
\usepackage[capitalize]{cleveref}
\usepackage{mathpartir}
\usepackage{amssymb}
\usepackage{stmaryrd} %
\usepackage{xspace}
\usepackage{latexsym}
\usepackage{latexsym}
\usepackage{ifthen}
\usepackage{color}
\usepackage[usenames,dvipsnames]{xcolor}
\usepackage{listings}
\usepackage{verbatim}
\usepackage[colorinlistoftodos]{todonotes} 
\usepackage{tikz}
\usetikzlibrary{positioning,shadows,arrows,calc,backgrounds,fit,shapes,snakes,shapes.multipart,decorations.pathreplacing,shapes.misc,patterns}
\usepackage{xspace}
\usepackage[T1]{fontenc}
\usepackage[scaled=.83]{beramono}
\usepackage{epigraph}
\usepackage{booktabs}
\usepackage{float}
\usepackage{etoolbox}
\usepackage{wrapfig}
\usepackage{pgfplots}
\usepackage{nameref}
\usepackage{scalerel}
\usepackage[framemethod=tikz]{mdframed}

\usepackage{bm}
\usepackage{tabularx}
\usepackage[square,sort,comma,numbers]{natbib}

\hypersetup{ pdfpagemode=UseOutlines, colorlinks=true, linkcolor=ForestGreen, citecolor=ForestGreen }

\theoremstyle{definition}
\newtheorem{example}{Example}
\newcommand{\mi}[1]{\ensuremath{\mathit{#1}}}
\newcommand{\mr}[1]{\ensuremath{\mathrm{#1}}}

\newcommand{\mtt}[1]{\ensuremath{\mathtt{#1}}}
\newcommand{\mf}[1]{\ensuremath{\mathbf{#1}}}

\newcommand{\ms}[1]{\ensuremath{\mathsf{#1}}}

\newcommand*{\QEDA}{\hfill\ensuremath{\blacksquare}}%

\AtEndEnvironment{problem}{\null\hfill\QEDA}
\AtEndEnvironment{example}{}%

\Crefname{lstlisting}{Listing}{Listings}
\Crefname{problem}{Problem}{Problems}

\Crefname{equation}{Rule}{Rules}

\newcommand{\compskel}[3]{\ensuremath{\bl{\left\llbracket \src{#1} \right\rrbracket^{#2}_{#3}}}}
\newcommand{\comp}[1]{\compskel{#1}{}{}}

\newcommand{\funname}[1]{\mtt{#1}}
\newcommand{\fun}[2]{\ensuremath{{\bl{\funname{#1}\left(#2\right)}}}\xspace}

\renewcommand{\S}[0]{\src{{S}}\xspace}
\newcommand{\T}[0]{\trg{{T}}\xspace}

\newcommand{\hole}[1]{\ensuremath{\left[#1\right]}}

\newcommand{\neutcol}[0]{black}
\newcommand{\stlccol}[0]{RoyalBlue}
\newcommand{\ulccol}[0]{RedOrange}
\newcommand{\commoncol}[0]{black}   

\newcommand{\col}[2]{\ensuremath{{\color{#1}{#2}}}}

\newcommand{\bl}[1]{\col{\neutcol }{#1}}
\newcommand{\com}[1]{\mi{\col{\commoncol }{#1}}}

\newcommand{\src}[1]{\ms{\col{\stlccol}{#1}}}
\newcommand{\srcu}[1]{{\color{\stlccol}\ensuremath{\mbox{\unboldmath{\(\mathsf{#1}\)}}}}}
\newcommand{\trgb}[1]{\mf{\bm{\col{\ulccol }{#1}}}}
\newcommand{\trg}[1]{{\mf{\col{\ulccol }{#1}}}}

\newcommand{\othercol}[0]{Emerald}
\newcommand{\oth}[1]{\mi{\col{\othercol }{#1}}}

\newcommand{\orcol}[0]{YellowOrange}
\newcommand{\con}[1]{\mtt{\col{\orcol }{#1}}}

\newcommand{\fifcol}[0]{CarnationPink}
\newcommand{\fif}[1]{{\col{\fifcol }{\underline{\ms{#1}}}}}

\newcommand{\cone}[1]{\src{1}}
\newcommand{\ctwo}[1]{\trg{1}}
\newcommand{\cthr}[1]{\oth{1}}
\newcommand{\cfou}[1]{\con{1}}
\newcommand{\cfiv}[1]{\fif{1}}

\newcounter{typerule}
\crefname{typerule}{rule}{rules}

\newcommand{\typeruleInt}[5]{%
	\def\thetyperule{#1}%
	\refstepcounter{typerule}%
	\label{tr:#4}%
  \ensuremath{\begin{array}{c}#5 \inference{#2}{#3}\end{array}} 
}
\newcommand{\typerule}[4]{%
  \typeruleInt{#1}{#2}{#3}{#4}{\textsf{\scriptsize ({#1})} \\      }
}

\pgfdeclarelayer{background}
\pgfdeclarelayer{veryback}
\pgfdeclarelayer{veryback2}
\pgfdeclarelayer{veryback3}
\pgfdeclarelayer{back2}
\pgfdeclarelayer{foreground}
\pgfdeclarelayer{foreground2}
\pgfdeclarelayer{foreground3}
\pgfdeclarelayer{foreground4}
\pgfsetlayers{veryback3,veryback2,veryback,background,back2,main,foreground,foreground2,foreground3,foreground4}

\newcommand{\link}[1]{\href{#1}{#1}}

\definecolor{mygreen}{rgb}{0,0.6,0}
\definecolor{mygray}{rgb}{0.5,0.5,0.5}
\definecolor{mymauve}{rgb}{0.58,0,0.82}

\lstdefinelanguage{Java} %
{morekeywords={abstract, all, and, as, assert, but, check, disj, else, exactly, extends, fact, for, fun, iden, if, iff, implies, in, Int, int, let, lone, module, no, none, not, one, open, or, part, pred, run, seq, set, sig, some, sum, then, univ, package, class, public, private, null, return, new, interface, extern, object, implements, System, static, super, try , catch, throw, throws, Unit, var, val, of, principal, trust},
sensitive=true,
keywordstyle=\bfseries\color{green!40!black},
commentstyle=\itshape\color{purple!40!black},
morecomment=[l][\small\itshape\color{purple!40!black}]{//},
identifierstyle=\color{blue},
stringstyle=\color{orange},
basicstyle=\small,
basicstyle={\small\ttfamily},
numbers=left,
numberstyle=\tiny\color{mygray},
tabsize=2,
numbersep=3pt,
breaklines=true,
lineskip=-2pt,
stepnumber=1,
captionpos=b,
breaklines=true,
breakatwhitespace=false,
showspaces=false,
showtabs=false,
float=!h,
columns=fullflexible,escapeinside={(*@}{@*)},
moredelim=**[is][\color{red!60}]{@}{@},
literate={->}{{$\to$}}1 {^}{{$\mspace{-3mu}\widehat{\quad}\mspace{-3mu}$}}1
{<}{$<$ }2 {>}{$>$ }2 {>=}{$\geq$ }2 {=<}{$\leq$ }2
{<:}{{$<\mspace{-3mu}:$}}2 {:>}{{$:\mspace{-3mu}>$}}2
{=>}{{$\Rightarrow$ }}2 {+}{$+$ }2 {++}{{$+\mspace{-8mu}+$ }}2
{<=>}{{$\Leftrightarrow$ }}2 {+}{$+$ }2 {++}{{$+\mspace{-8mu}+$ }}2
{\~}{{$\mspace{-3mu}\widetilde{\quad}\mspace{-3mu}$}}1
{!=}{$\neq$ }2 {*}{${}^{\ast}$}1 %
{\#}{$\#$}1
}
\DeclareMathOperator\ceq{\ensuremath{\mathrel{\simeq}}}

\DeclareMathOperator\ceqt{\trgb{\ceq}}

\theoremstyle{definition}

\Crefname{corollary}{Corollary}{Corollaries}
\Crefname{informal}{Definition}{Definition}
\Crefname{assumption}{Assumption}{Assumptions}
\crefname{assumption}{Assumption}{Assumptions}
\Crefname{property}{Property}{Properties}
\crefname{property}{Property}{Properties}

\newcommand{\lamt}[2]{\ensuremath{\trgb{\lambda} #1\ldotp #2}}

\newcommand{\ifte}[3]{\ensuremath{{if}~#1~{then}~#2~{else}~#3}}

\newcommand{\letin}[3]{{let}~#1=#2~{in}~#3}
\newcommand{\letins}[3]{\src{let}~#1\src{=}#2~\src{in}~#3}

\AtEndEnvironment{example}{\null\hfill$\boxdot$}

\makeatletter
\xdef\@thefnmark{\@empty}

\makeatother

\newcounter{hps}
\crefname{hps}{}{}

\newcommand{\xto}[1]{\ensuremath{~\mathrel{\xrightarrow{~#1~}}~}}

\def\teqaux#1{\vcenter{\hbox{\ooalign{\hfil
       \raise6pt \hbox{\scriptsize{T}}\hfil\cr\hfil
       $=$}}}}

\title{\vspace*{-2cm}
	Why Should Anyone use Colours?
	\\
	{\large
		or, Syntax Highlighting Beyond Code Snippets
	}
}
\author{
	Marco Patrignani$^{\dagger,\ddagger}$
	\\
	\textit{
		$^\dagger$~Stanford University 
	}
	\\
	\textit{
		$^\ddagger$~CISPA Helmholtz Centre for Information Security 
	}
	\\
	\texttt{
		email at: \link{https://squera.github.io/}
	}
}
\date{}

\begin{document}
\maketitle

\begin{abstract}
	Syntax highlighting in the form of colours and font diversification, is an excellent tool to provide clarity, concision and correctness to writings.
	Unfortunately, this practice is not widely adopted, which results in often hard-to-parse papers.
	The reasons for this lack of adoption is that researchers often struggle to embrace new technologies, piling up unconvincing motivations.
	
	This paper argues against such motivations and justifies the usage of syntax highlighting so that it can become a new standard for dissemination of clearer and more understandable research.
	Moreover, this paper reports on the criticism grounded on the shortcomings of using syntax highlighting in \LaTeX\ and suggests remedies to that.

	We believe this paper can be used as a guide to using syntax highlighting as well as a reference to counter unconvincing motivations against it.
\end{abstract}

\section{Introduction}\label{sec:intro}

Syntax highlighting (SH), in the form of colours and font diversification, is an excellent tool to provide clarity, concision and correctness to writings.
Generally, in most webpages hypertext links are coloured in blue or underlined (whereas normal text is black and not underlined).
That syntax highlighting is done for the sake of clarity: writers put effort in devising links that readers can follow to further their knowledge.
More specifically, many researchers are already used to seeing a form of syntax highlighting, for example, in code snippets.
There, keywords, identifiers, literal and comments are all given different colours and different fonts, as the familiar code snippet below points out.
\begin{lstlisting}
	public static void main (int* args) { return 0; } // comment
\end{lstlisting}

However, this is not done widely, and some areas of research suffer from this lack of practice.
Let us now consider and example from computer science literature, specifically, when describing a compiler from a language $L_s$ to another language $L_t$.%
\footnote{%
	Several other examples exist, for example comparing different systems, or models, or languages for expressive power, one often has to indeed first define both systems.
	We treat those examples later in the paper.
}
\begin{quote}\it
	1) A compiler \comp{\cdot} is a function from programs $P$ of language $S$ to programs $P$ of language $T$, which we denote as \comp{\com{P}}.

	2) The behaviour of a program $P$ is a set of traces defined as $\fun{B}{P}$.

	3) For a correct compiler, the behaviour of a compiled program \comp{\com{P}} is a subset of the behaviour of the original program $P$, i.e., $\fun{B}{\comp{\com{P}}}\subseteq\fun{B}{P}$.
\end{quote}

This example already shows some shortcomings of not using any form of syntax highlighting.
In point 1, the same meta-variable $P$ denotes both elements of a source language $L_s$ and elements of a target language $L_t$, which can be very different between each other.
A simple solution is that to annotate said meta-variable $P$ with a subscript $S$ or $T$ to attribute it to the right language.
However, this has several drawbacks, as it often ends up polluting equations and formulas with additional subscripts.

A much more compact solution would be to just write all elements of $S$ and of $T$ in two different colours and fonts.
For example, we can write all elements of $S$ in a \src{blue}, \src{sans}-\src{serif} font and all elements of $T$ in a \trg{red}, \trg{bold} one.%
\footnote{%
	The choice of this specific syntax highlighting scheme is motivated later in this paper.
}
The result is much more concise and clear.
\begin{quote}\it
	1) A compiler \comp{\cdot} is a function from programs $\src{P}$ of language $\S$ to programs $\trg{P}$ of language $\T$, which we denote as \comp{\src{P}}.
\end{quote}

Another crafty shortcoming of not using any form of SH can be seen in point 3.
There, the equation $\fun{B}{\comp{\com{P}}}\subseteq\fun{B}{P}$ is seemingly correct.
However, if we try to colour it, we realise something is afoot: $\fun{\trg{B}}{\comp{\src{P}}}\subseteq\fun{\src{B}}{\src{P}}$, we are comparing sets whose elements are \emph{different}.

Of course without any SH this is rather tricky to notice, while with basic highlighting such as colours it is not.
Again, some people may advocate resorting to subscripts, but this complicates things.
In fact one first has to generalise point 2 and the notation for behaviours as follows.
\begin{quote}\it
	2-b) The behaviour of a program $P_S$ of language $S$ is a set of traces defined as $\fun{B_S}{P_S}$.
\end{quote}
Then, one can write point 3 with subscripts, as follows: 
\begin{align*}
	\fun{B_T}{\comp{\com{P_S}}}\subseteq\fun{B_S}{P_S}
\end{align*}
Again, this solution is not concise and it does not scale if the formula at hand has several elements from the different languages.
This can be seen in the next example that compares behaviours of two programs linked via operator $+$ (which can vary wildly between \S and \T).%
\footnote{
	This example also highlights that subscripts ``consume'' another common place for identifiers, if we had a matrix of programs $P_i^j$, where to put the language annotation would not be obvious.
}
For clarity we report the syntax-highlighted version too.
\begin{align*}
	\fun{B_T}{\comp{\com{P_S^1}}+_T\comp{\com{P_S^2}}}\subseteq&\ \fun{B_S}{P_S^1+_SP_S^2}
	\\
	\fun{\trg{B}}{\trg{\comp{\src{P_1}}\trgb{+}\comp{\src{P_2}}}}\subseteq&\ \fun{\src{B}}{\src{P_1+P_2}}
\end{align*}

Unfortunately, the usage of SH is not widespread and often researchers struggle to embrace these technologies, piling up unconvincing motivations.
The most notorious ones are that colours are obscure to many reviewers (and we all know that baffling reviewers is to be avoided), that they will be hard to discern in printouts and that colourblind people will be hindered by them.

\paragraph{Contributions \& Outline}
This paper meticulously argues against such motivations and justifies the usage of SH (in the form of colours and font diversification) so that it can become a new standard for dissemination of clearer and more understandable research.

To motivate the use of SH, \Cref{sec:pros} argues in more depth about how to use colours and font diversification to improve clarity, concision and correctness.
Then, \Cref{sec:crit} presents often-voiced criticism about the usage of SH and argues against it.
Finally, \Cref{sec:short} discusses the main shortcomings of using SH in the form of colours and font diversification with \LaTeX\ and describes solutions to said shortcomings.

This work-in-progress is not yet complete, please forward any recommendation on SH, any criticism to be rebuked, and any \LaTeX\ hack to the author.

\section{Why Bothering}\label{sec:pros}
Throughout this section we will provide examples of when to use SH in the form of colours and font differentiation to increase clarity, conciseness and correctness of papers.
The following examples (\Cref{ex:ty-vs-syn,ex:diff-sys,ex:multilang}) discuss respectively the case of extending a system, of identifying different parts of a system and of identifying different systems.

\begin{example}[Additions to a system]\label{ex:ty-vs-syn}
When presenting an enrichment to a system, be it in terms of a program logic, a type system or a new semantics, it is often useful to highlight said enrichment to the reader.

In the case of program logics, this has the distinct benefit of clearly demarcating what must be supplied by possibly different entities, namely the program and the annotations.
\begin{center}
	\typerule{Frame}{
		\src{\{p\}} ~c~ \src{\{q\}}
	}{
		\src{\{p*r\}} ~c~ \src{\{q*r\}}
	}{}
	\typerule{Assignment}{}{
		\src{\{\ell\mapsto u\}} ~\ell:=v~ \src{\{\ell\mapsto v\}}
	}{}
\end{center}

In the case of type systems, this has already been pointed out to immediately highlight the Curry-Howard isomorphism~\citep{Wadler:2007:GI:1235896.1236119}.
In fact, by erasing the appropriate parts of the type-system, a logic appears~\citep{Bernardy:2013:TC:2544174.2500577}, and one can see the erasure simply by putting on glasses of the same colour as the syntax (blue in the case of the snippet below).
\begin{center}
	\typerule{Type-Lam}{
		\src{f:}A\to B & \src{v:}A
	}{
		\src{f~v:} B
	}{}
	\typerule{Modus Ponens}{A\to B & A}{ B}{}
\end{center}

Finally, consider a language and its operational semantics (indicated as \com{\xto{}}) being extended with another semantics with labels (indicated as \trg{\xto{l}}), that relies on the former.
In order to highlight where these dependencies happen we may want to colour semantics rules of the latter where this dependency is.
\begin{center}
	\typerule{Connecting Step}{
		\com{e \xto{} v}
	}{
		\com{e \trgb{\xto{\mr{\trg{l}}}} v}
	}{}
\end{center}

Moreover, we may want to state something about the semantics, e.g., that a term reducing according to the first also reduces according to the latter.
\begin{align*}
	\com{t \xto{} v} \Rightarrow \com{t \trgb{\xto{}} v}
\end{align*}
In this case SH helps ensuring that the reader knows that it is the right hand side of the equation that is concerned with the latter semantics.
Crucially, this is achieved without polluting the notation with unpleasant and confusing superscripts and subscripts, which may be used to denote other useful information (e.g., number of steps).
\end{example}

\begin{example}[Different parts of a system]\label{ex:diff-sys}
Another use case for SH comes from the world of compilation, as presented in \Cref{sec:intro}~\citep{mfac}.
Consider a modular compiler, i.e., one that operates on components and then links together its output to produce a larger program.
If we want to reason about its correctness we may want to express something like the following:
\begin{align*}
	\fun{\trg{B}}{\trg{\comp{C_1} \trgb{+} \cdots \trgb{+} \comp{C_n}}} \subseteq \fun{\src{B}}{\src{C_1 + \cdots + C_n}}
\end{align*}
Here, colours provide a crucial highlight, namely that there is a part of the two systems, which we write as $+$ that is not the same between the two systems.
In fact, $+$ may be the indication of a linker, and the usage of SH clarifies that there are \emph{different} kinds of linkers being employed in the two languages.

Abstracting away from details of the system at hand, we may be using the same notation to reason about the meta-system, e.g., to reason about equality of programs of the system.
This is, for example, the case for the statement of fully abstract compilation~\citep{DBLP:conf/icalp/Abadi98}:%
\footnote{
	Fortunately, many articles on related subjects use syntax highlighting~\citep{rhc,NewBA16,mfac,PatrignaniG18,Patterson:2017:FRM:3140587.3062347,10.1007/978-3-319-89366-2_8}.
}
\begin{align*}
	\forall\src{P_1},\src{P_2}. \src{P_1\ceq P_2}\iff \trg{\comp{P_1}\ceqt\comp{P_2}}
\end{align*}
In this case, the relation $\ceq$ that we use to relate programs can vary between the two languages, and the SH points that out nicely to the reader.
\end{example}

\begin{example}[Different systems interacting with each other]\label{ex:multilang}
Finally, we may be interested in capturing how different systems interact with each other, for example to study the foreign-function interface (FFI) of said systems.
This is often called a multilanguage system and it is indeed a scenario where colours have successfully been used to achieve clarity~\citep{Matthews:2007:OSM:1190215.1190220,Patterson:2017:FRM:3140587.3062347,Perconti:2014:VOC:2961744.2961759}.
In a multilanguage system where languages \src{I} and \trg{F} are calling each other, the FFI from the former to the latter is written as \com{IF} and the reverse is written as \com{FI}.
Thus, one can write snippets such as the following one, where some \src{imperative} \src{code} calls a \trg{functional} \trg{snippet}.
\begin{align*}
	\src{
		unsafe(x)\{
			\ifte{x}{ \left(\com{IF}\trg{\left(\lamt{y}{y\trgb{==}2}\right)\left(\com{FI}\src{x}\right)}\right) }{ skip }
		\}
	}
\end{align*}
The explicit boundaries as well as the SH helps the reader understanding why suddenly there is some functional code in the middle of imperative one.
\end{example}

\section{Criticism to Syntax Highlighting}\label{sec:crit}
This section dissects the most often-heard criticism to the adoption of SH, i.e., that colourblind people are hindered by it (\Cref{sec:cb}) and black and white printers do not render SH nicely (\Cref{src:pr}).
This section also discusses the hardest criticism coming from people that simply are not used to SH (\Cref{sec:notused}).

\subsection{Colourblind}\label{sec:cb}
Colourblindness is a serious genetic condition that affects the 8\% of male population worldwide and a smaller fraction of female population, as the gene responsible for colourblindness is found in the X chromosome.
Thus, there is a significant chance that colourblind people will attend your talks and read your papers and this is what many researchers point out to people using colours.
Sadly, this is also where most researchers stop, not noticing that effective measures can be taken to ensure that colourblind people also enjoy coloured papers.
To ensure colourblind people can make out what is going on when colours are used, two things can be done:
\begin{enumerate}
	\item use font diversification;
	\item use colours that are still distinguishable for colourblind people.
\end{enumerate}
Concerning point 1, that is why throughout this paper we use \src{sans}-\src{serif} and \trg{bold} fonts.
Concerning point 2, there exist tools such as Sim Daltonism%
	\footnote{ 
		Available for MAC OS X at: \link{https://michelf.ca/projects/sim-daltonism/}
	} 
that allows us to conduct an analysis on what colours can still be perceived by people with different kinds of colourblindness.
The combination \src{blue} versus \trg{red} is, for example, still discernible from \emph{all} kind of colourblind people (who can still perceive colours).
So it is clear: using a \src{blue}, \src{sans}-\src{serif} font and a \trg{red}, \trg{bold} one voids any claim that the paper is inaccessible to colourblind people.

A more advanced concern pops up now: what if there are more than two systems in our paper?
What is the best combination for a third system, and a fourth one and so on?
Unfortunately, there are only so many colour and font combination available.
Luckily writing papers with more than two systems is already challenging, more than three is very rare and more than four is seemingly unheard of.
The list below summarises SH per number of systems, while the table below shows the SH in action on different kinds of text: characters, words, greek letters and math.
The colour between brackets is the precise \LaTeX colour that each line uses.
\begin{enumerate}
	\item \src{blue} (\src{RoyalBlue}), \src{sans}-\src{serif};
	\item \trg{red} (\trg{RedOrange}), \trg{bold}, \trg{with}-\trg{serif};
	\item \oth{emerald} (\oth{Emerald}), \oth{italic}, \oth{with}-\oth{serif};
	\item \con{orange} (\con{YellowOrange}), \con{monospaced}, \con{with}-\con{serif};
	\item \fif{pink} (\fif{CarnationPink}), \fif{underlined}, \fif{sans}-\fif{serif}.
\end{enumerate}

\begin{center}
\begin{tabular}{| l | c | c | c | c | c |}
	\hline
		&
			\textbf{First}
		&
			\textbf{Second}
		&
			\textbf{Third}
		&
			\textbf{Fourth}
		&
			\textbf{Fifth}
	\\
	\hline
	\hline
	\textbf{Character}
		&
			\src{a}~\src{z}~\src{g}
			&
			\trg{a}~\trg{z}~\trg{g}
				&
			\oth{a}~\oth{z}~\oth{g}
					&
			\con{a}~\con{z}~\con{g}
						&
			\fif{a}~\fif{z}~\fif{g}
	\\
	\hline
	\textbf{Word}
		&
			\src{gazebo}
			&
			\trg{gazebo}
				&
			\oth{gazebo}
					&
			\con{gazebo}
						&
			\fif{gazebo}
	\\
	\hline
	\textbf{Greek}
		&
			\src{\alpha}~\src{\Pi}
			&
				\trgb{\alpha}~\trgb{\Pi}
				&
					\oth{\alpha}~\oth{\Pi}
					&
					\con{\alpha}~\con{\Pi}
						&
					\fif{\alpha}~\fif{\Pi}
	\\
	\hline
	\textbf{Symbol}
		&
			\src{+}~\src{\ceq}
			&
			\trgb{+}~\trg{\ceqt}
				&
			\oth{+}~\oth{\ceq}
					&
			\con{+}~\con{\ceq}
						&
			\fif{+~\ceq}
	\\
	\hline
\end{tabular}
\end{center}

\medskip

In order to understand how colourblind people perceive the use of SH, we provide the list of images below.
Those images are obtained through the use of Sim-Daltonism, an application that replicates different forms of colourblindness.  
The images show how people with different visual impairment (in the title of the frames) see the table above.
This list serves to inform readers without visual impairment that the choice of colours and of fonts still result in visually distinct syntax highlighting that colourblind people can benefit from.

\noindent\includegraphics[width=1.\textwidth]{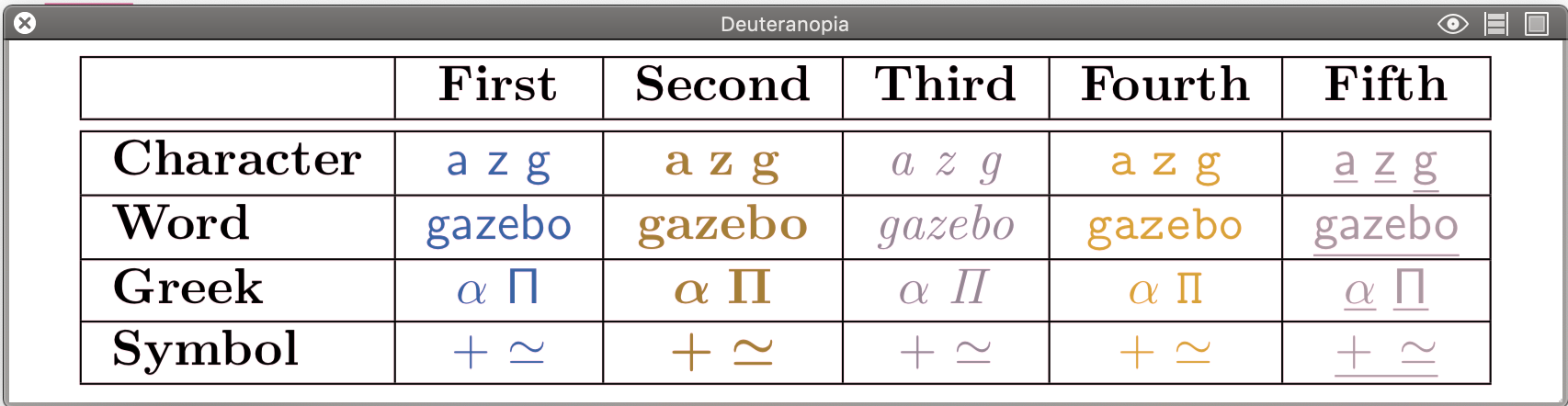}

\noindent\includegraphics[width=1.\textwidth]{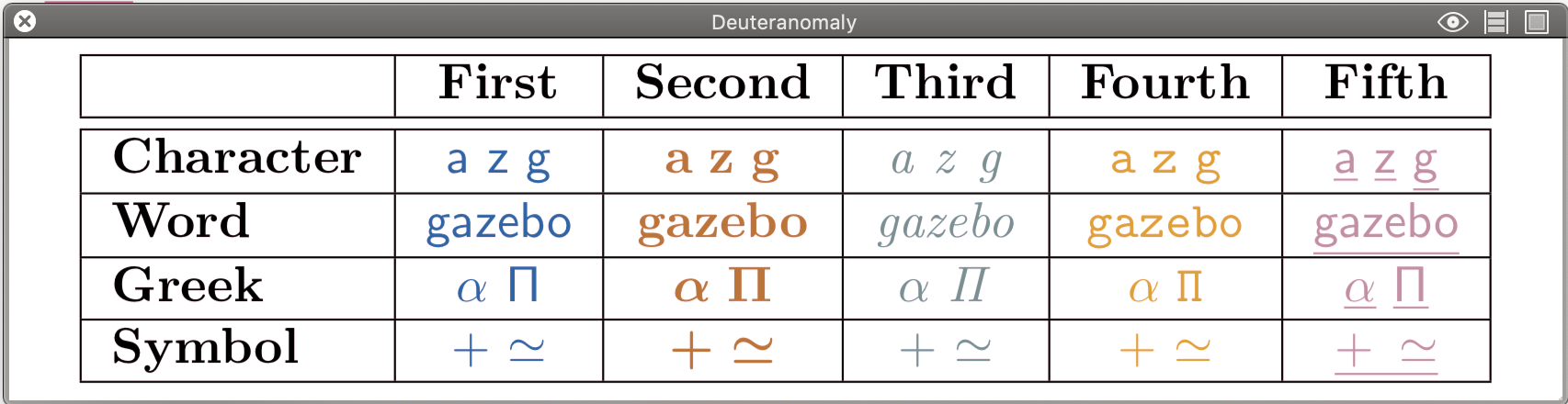}

\noindent\includegraphics[width=1.\textwidth]{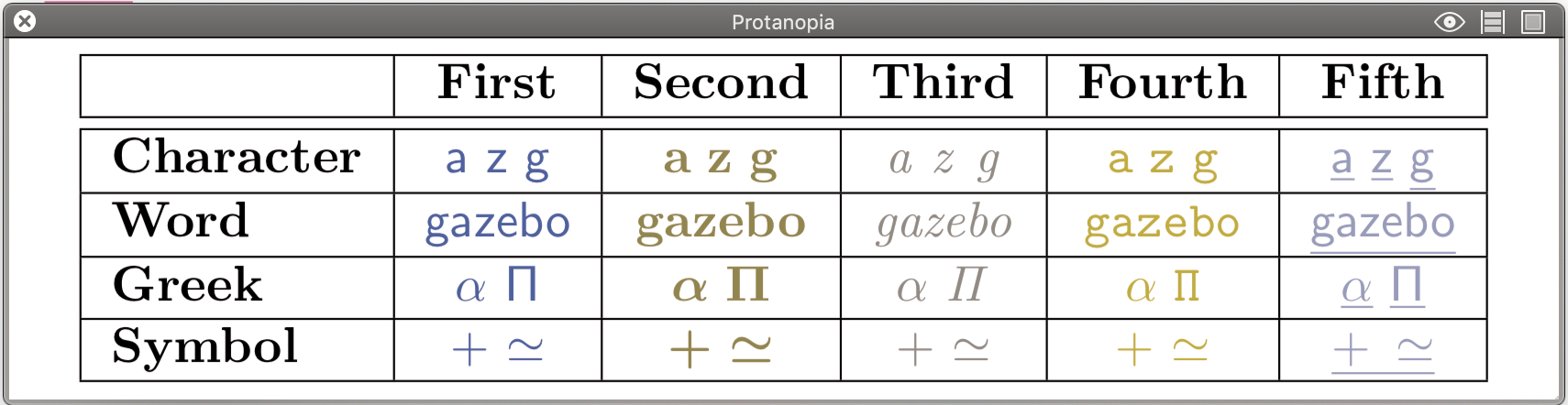}

\noindent\includegraphics[width=1.\textwidth]{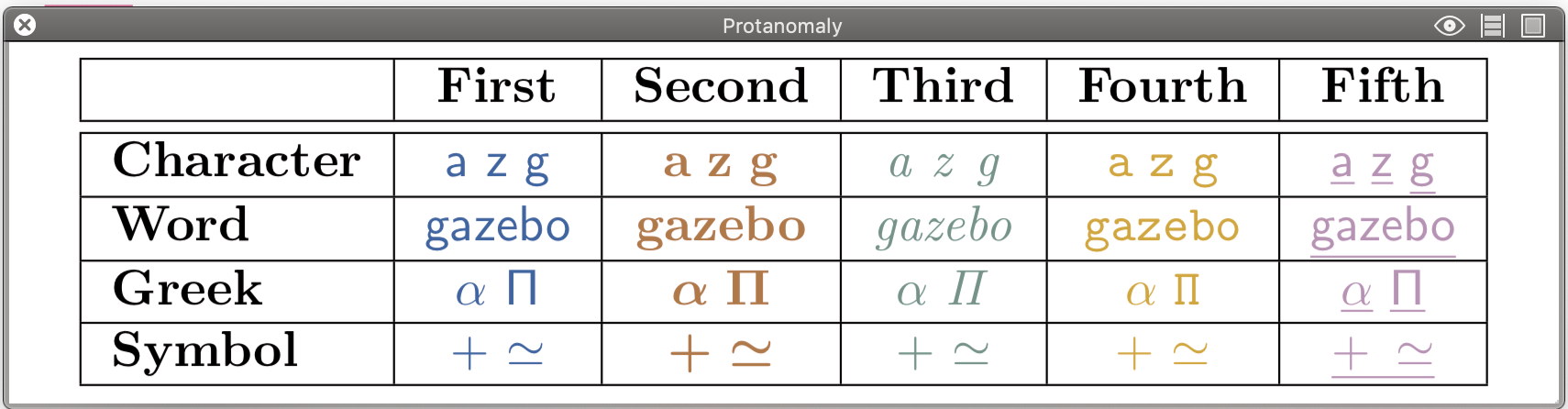}

\noindent\includegraphics[width=1.\textwidth]{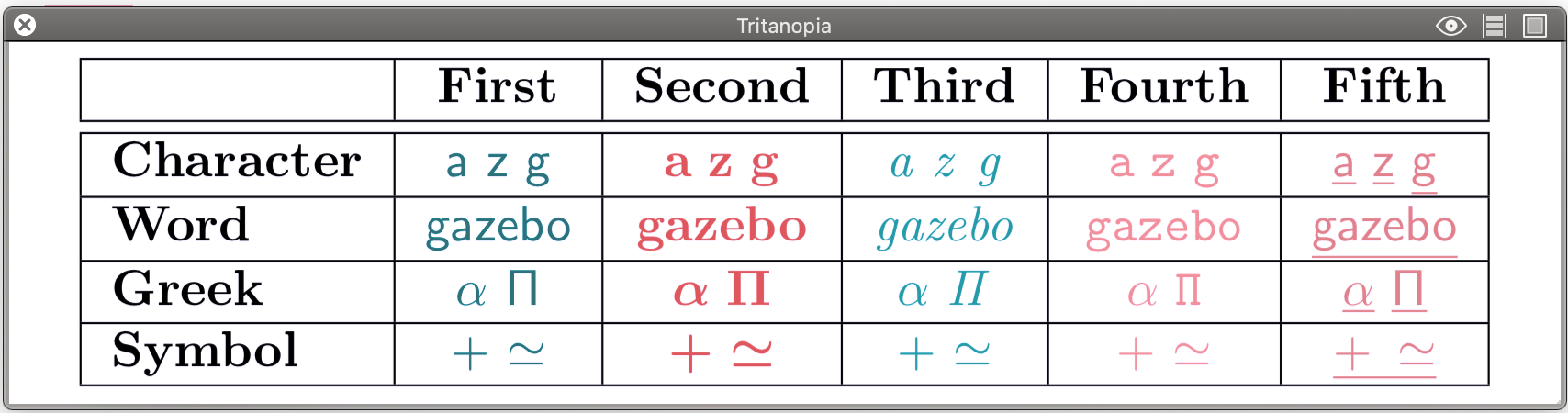}

\noindent\includegraphics[width=1.\textwidth]{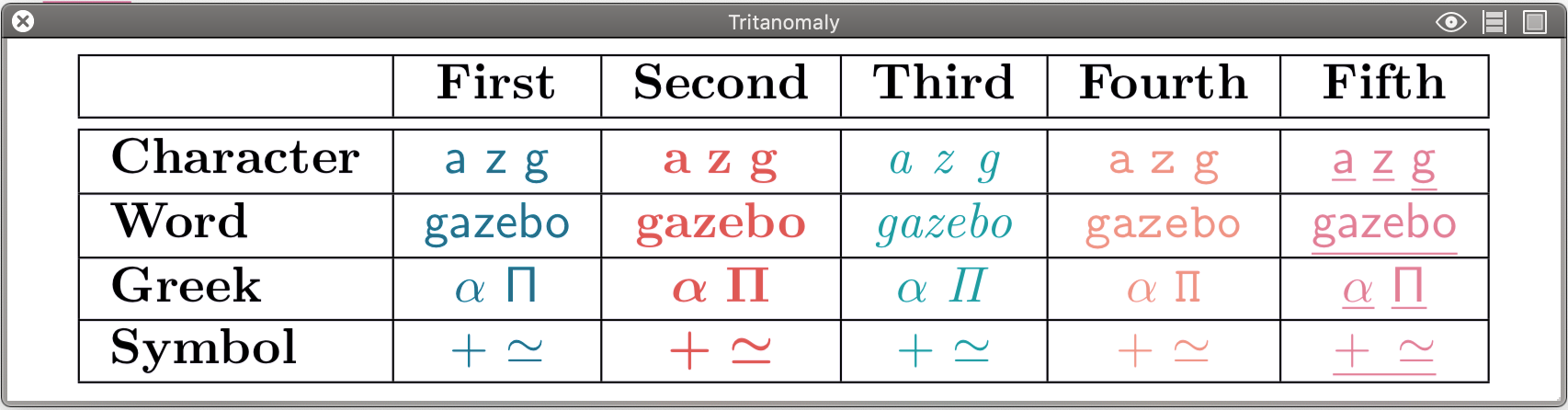}

\noindent\includegraphics[width=1.\textwidth]{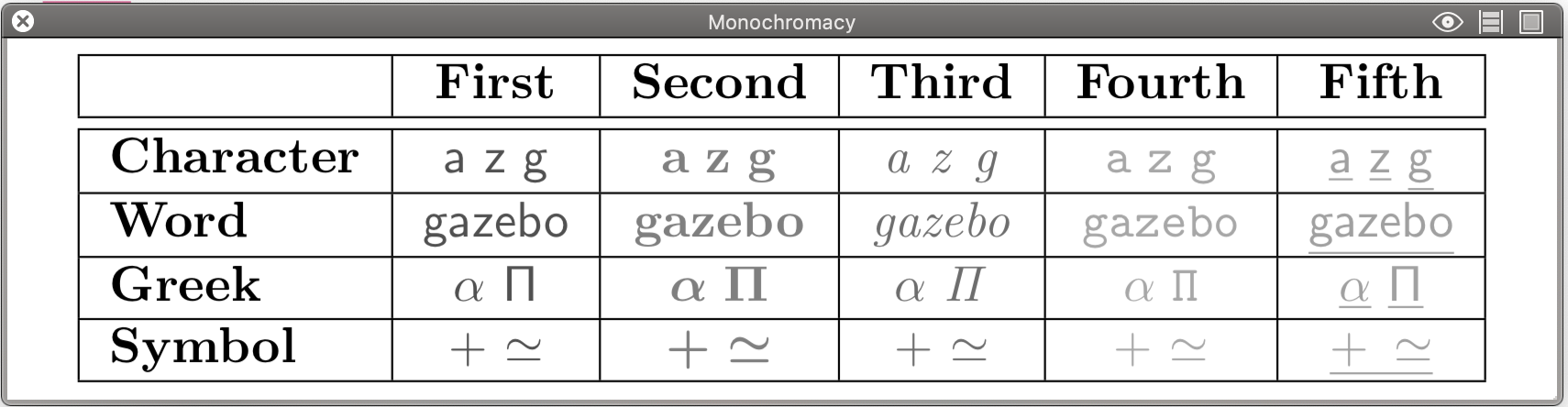}

\noindent\includegraphics[width=1.\textwidth]{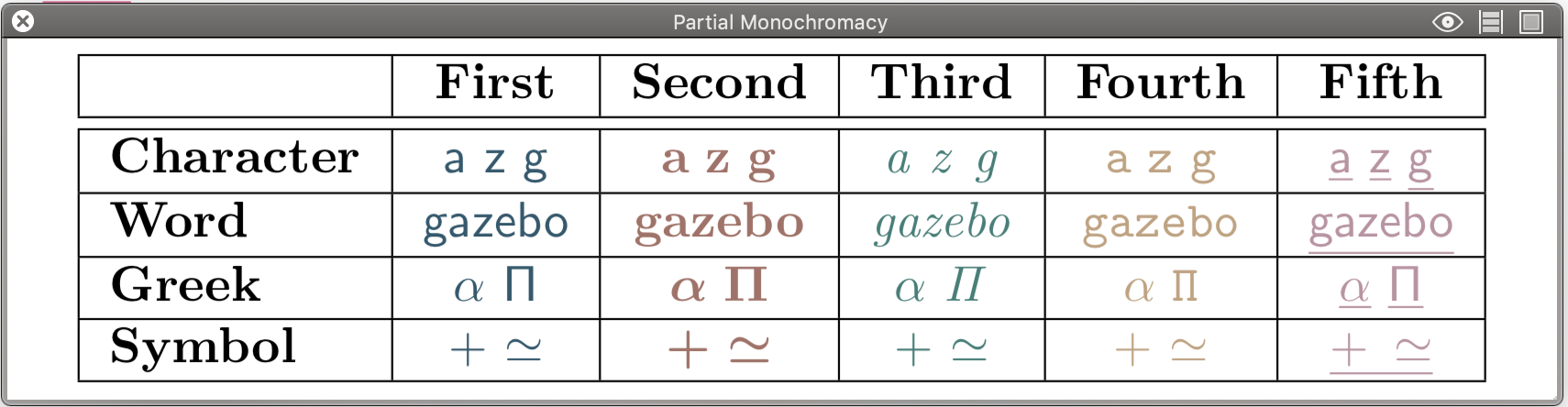}

\subsection{Black and White Printouts}\label{src:pr}
Some researchers argue that when printing on black and white settings, there is not much difference between the shades of grey to effectively distinguish the colours.
This seems a rather short-lived comment for several reasons.

First, font diversification carries out to black and white printouts, so already the font gives meaning to distinguish elements with similar shades of grey.
This is particularly visible in the second-to-last picture from the list above --- labelled Monochromacy --- which replicates how people that can only distinguish shades of grey see the syntax highlighting we propose.
Such a colouring is exactly what a black and white printer also produces.

Second, and most importantly, paper printouts are a dying medium and we should encourage its death.
With the advent of modern technologies such as hi-resolution tablets, it seems an environmentally unhealthy choice to keep printing papers that will be read once, and then recycled (which is still a costly, albeit noble, process).

\subsection{Accepting a New Notation}\label{sec:notused}
The hardest criticism to using and employing colours is, often, simply that of a mental barrier.
This is still a rather new notation, with a slowly-growing and still small group of adopters.
Although it takes a little effort to get into the colour-usage mentality, many are simply unwilling to take this step.

The best way to convert these researchers is to adopt this notation ourselves, use it and make it mainstream, showing clear, concise, elegant work.
Then they will realise it takes a little step to understand (and use) colours, and they have already done that step by looking at our work that does already use colour.

\section{Colours and \LaTeX}\label{sec:short}

Using colours to typeset papers in \LaTeX\ can be rather complex and lead to numerous concerns.
Fortunately, this section presents some notorious issues when compiling coloured code and solutions to them.

\subsection{Colour Choice}
A first concern is that the standard \LaTeX\ colours are not very pleasant: they are very bright, almost too vibrant colours and a fully-coloured page would hurt the eyes.
The best solution to this is to use the \texttt{xcolor} package, which provides \src{RoyalBlue} replacing the {\color{blue}{latex\ blue}}, \trg{RedOrange} replacing {\color{red}{latex\ red}} and {\color{ForestGreen}{ForestGreen}} replacing {\color{green}{latex\ green}} among many other more pleasant hues such as {\color{CarnationPink}{CarnationPink}} for pink text.

So, do not forget to include:

\texttt{\textbackslash usepackage[usenames,dvipsnames]\{xcolor\}}

Some templates (like the ones used by PACMPL currently) already import \texttt{xcolor}, so you cannot re-import it and define your own options.
Instead, the following can be written before the \texttt{\textbackslash documentclass} to load the correct options.

\texttt{\textbackslash PassOptionsToPackage\{names,dvipsnames\}\{xcolor\}}

\subsection{Colouring Macros}
Once the \texttt{xcolor} package is loaded, we can define commands that perform syntax highlighting.
To keep in line with the compilation examples from \Cref{sec:intro}, we define two commands for typesetting blue source and red target elements:

\texttt{\textbackslash newcommand\{\textbackslash src\}\hole{1}\{\textbackslash color\{RoyalBlue\}\{\textbackslash mathsf\{\#1\}\}\}}

\texttt{\textbackslash newcommand\{\textbackslash trg\}\hole{1}\{\textbackslash color\{RedOrange\}\{\textbackslash mathbf\{\#1\}\}\}}

A pre-made structuring of macros that perform syntax highlighting can be found in the sources of this paper on arXiv or at the following link: 

\noindent\link{http://squera.github.io/misc/cmds.tex}.

\subsection{Colours in Titles and Headings}
Sometimes having a \texttt{\textbackslash src\{\src{.}\}} (or an analogous SH command) inside a section title will make \LaTeX\ not build.
In that case one has to redefine the colour name.
For example, for a colour called \emph{RoyalBlue} one has to write the following to ensure \LaTeX will build.

\texttt{\textbackslash colorlet\{ROYALBLUE\}\{RoyalBlue\}}

\subsection{Math, Symbols, Alignments and Colours}
It is indeed complicated to ensure macros do not break when used within equations and aligned environments in \LaTeX. 
Here are few tips to ensure one can write modular macros that will build in most settings.
\begin{enumerate}
	\item Define symbols with their colouring.

	For example, if you need to define a relation such as $\approx$ (\texttt{\textbackslash approx}) in different colours, it will come in handy to have \src{\approx} (\texttt{\textbackslash srcapprox}) as a single new command.

	\item Define bold symbols.

	Bold math is very complex to get right and robust, since it relies on the brittle \texttt{bm} package.
	As such, it is recommended to have a command for typesetting bold symbols in the second system:

	\texttt{\textbackslash newcommand\{\textbackslash trgbold\}\hole{1}\{\textbackslash color\{RedOrange\}\{\textbackslash mathbf\{\textbackslash bm\{1\}\}\}\}}

	This way, when we need to write just text, e.g., \trg{cat} we can write \texttt{\textbackslash trg\{cat\}}, but when we are writing a symbol as \trgb{\approx}, we instead write \texttt{\textbackslash trgbold\{\textbackslash approx\}}.

	\begin{enumerate}
		\item Nesting commands.

		In case you want to be able to nest commands where one may use \texttt{\textbackslash bm}, other non-bold commands should contain \texttt{\textbackslash unboldmath}, as follows:

		\texttt{\textbackslash newcommand\{\textbackslash srcu\}\hole{1}\{\textbackslash color\{RoyalBlue\}\{\textbackslash ensuremath\{\textbackslash unboldmath \textbackslash(\{\textbackslash mathsf\{\#1\}\}\textbackslash)\}\}\} }

		This way you can carelessly write a top-level \texttt{\textbackslash trgbold} and supply \texttt{\textbackslash srcu} text as part of the argument.
		For example \trgb{ \srcu{A} \approx \srcu{B}} is obtained by just writing 
		\texttt{ \textbackslash trgbold\{\textbackslash srcu\{A\} \textbackslash approx \textbackslash srcu\{B\}\} }

		To ensure no \LaTeX\ errors pop up, be sure to include this in your preamble \emph{before} loading the \texttt{bm} package:%
		\footnote{
			Cfr: \link{https://tex.stackexchange.com/questions/3676/too-many-math-alphabets-error}
		}
		
		\texttt{\textbackslash newcommand\textbackslash hmmax{0}}

		\texttt{\textbackslash newcommand\textbackslash bmmax{0}}
	\end{enumerate}

	\item Define black symbols.

	The same idea above applies to black symbols, e.g., having $\comp{\cdot}$ as a single new command.
	If we have a command \texttt{\textbackslash blk} to write black text, we can define \comp{\cdot} (comp) as follows:

	\texttt{\textbackslash newcommand\{\textbackslash blk\}\hole{1}\{\textbackslash color\{black\}\{\#1\}\}}
	
	\texttt{\textbackslash newcommand\{\textbackslash comp\}\hole{1}\{\textbackslash blk\{\textbackslash left\textbackslash llbracket\textbackslash src\{\#1\}\textbackslash right\textbackslash rrbracket\}\} }
	 
	That way one does not have to worry when embedding such a symbol in a larger coloured text such as \trg{\comp{C}\trgb{+}C}, which we can simply write as:

	\texttt{
		\textbackslash trg\{\textbackslash comp\{C\} \textbackslash trgbold\{+\} C\}
	}
	
	\item No wrapping of parameters in SH.

	Let us assume we have a macro to typeset a let-in construct that takes the three parameters of the letin and produces the following: \com{\letin{x}{v}{e}}.
	We may define such a macro as follows:

	\texttt{\textbackslash newcommand\{\textbackslash letin\}[3]\{let$\sim$\#1=\#2$\sim$in$\sim$\#3\}}

	In order for it to appear correctly coloured, we have to provide a coloured variant as follows:

	\texttt{\textbackslash newcommand\{\textbackslash srcletin\}[3]\{\textbackslash src\{let\}$\sim$\#1\textbackslash src\{=\}\#2$\sim$\textbackslash src\{in\}$\sim$\#3\}}

	It is important not to wrap the arguments in \texttt{\textbackslash src\{$\cdot$\}} otherwise when used within an aligned environment, things may break.
	In fact, if we were to write 

	\texttt{\textbackslash newcommand\{\textbackslash wrongsrcletin\}[3]\{\textbackslash src\{let$\sim$\#1\}\textbackslash src\{=\#2\}$\sim$\textbackslash src\{in$\sim$\#3\}\}}

	where the \texttt{\textbackslash src\{$\cdot$\}} commands encompasses the arguments too, those arguments will now by part of a sub-group.
	Thus we would not be able to place alignment markers such as ``\&'' inside them, as in the example below.
	There, the third argument is ``\texttt{\textbackslash\textbackslash~\&\textbackslash~x}'' and this ensures a proper alignment of subparts of the command (as can be seen below). 

	\begin{minipage}{.45\textwidth}
	\begin{align*}
		&
		\letins{\src{x}}{\src{v}}{
			\\
			&\
			\src{x}
		}
	\end{align*}
	\end{minipage}\hfill
	\begin{minipage}{.45\textwidth}
	\begin{lstlisting}[mathescape]
		\begin{align*}
		&
		\srcletin{\src{x}}{\src{v}}{
			\\
			&\
			\src{x}
		}
		\end{align*}
	\end{lstlisting} 
	\end{minipage}

	Unfortunately, not all that glitters is gold, which means we must wrap the contents of the arguments that are not alignment markers in \texttt{\textbackslash src\{$\cdot$\}}.
	
\end{enumerate}

\section{Conclusion}
This paper argued in favour of using syntax highlighting in the form of colours and font diversification to provide more clarity, concision and correctness to research papers.
This paper discussed the pros of using colours and font diversification, how to prevent poorly-grounded criticism against it and how to overcome technical difficulties when typesetting coloured work with \LaTeX.

\bigskip
{\small\textbf{Acknowledgement:}
	This work has benefitted from discussions with the following people as well as from collaborating with a subset of them (and from the many reviewer comments we had for said collaborations):
	Carmine Abate, Amal Ahmed (my first reference for colour usage), Roberto Blanco, William Bowman, Dave Clarke, Stefan Ciobaca (thanks for pointing out the usage of \texttt{\textbackslash unboldmath}), Akram El-Korashi, Dominique Devriese, Deepak Garg, Catalin Hritcu, Adriaan Larmuseau, Max New, Daniel Patterson, Frank Piessens, Jeremy Thibault.
}

\newpage
\bibliography{./bibliobib.bib}

\end{document}